\documentclass[%
 reprint,
superscriptaddress,
 amsmath,amssymb,
 aps,
]{revtex4-1}

\usepackage{graphicx}
\usepackage{dcolumn}
\usepackage{bm}
\usepackage{hyperref}
\usepackage[dvipsnames]{xcolor}

\usepackage{subfiles}
\def \bnzo{BaNd$_2$ZnO$_5$}
\def \lnzo{BaLa$_2$ZnO$_5$}
\def \scbo{SrCu$_2$(BO$_3$)$_2$}
\usepackage{physics}  

\begin{document}

\preprint{APS/123-QED}

\title{Magnetic properties of the Shastry-Sutherland lattice material BaNd$_2$ZnO$_5$}
\thanks{This manuscript has been authored by UT-Battelle, LLC under Contract No. DE-AC05-00OR22725 with the U.S. Department of Energy.  The United States Government retains and the publisher, by accepting the article for publication, acknowledges that the United States Government retains a non-exclusive, paid-up, irrevocable, world-wide license to publish or reproduce the published form of this manuscript, or allow others to do so, for United States Government purposes.  The Department of Energy will provide public access to these results of federally sponsored research in accordance with the DOE Public Access Plan (http://energy.gov/downloads/doe-public-access-plan).}%

\author{Yuto Ishii}
\thanks{These authors contributed equally}
\affiliation{Department of Physics, Hokkaido University, North 10 West 8, Kita-ku, Sapporo 060-0810, Japan}
\affiliation{International Center for Materials Nanoarchitectonics (WPI-MANA), National Institute for Materials Science, 1-1 Namiki, Tsukuba, Ibaraki 305-0044, Japan}
 
\author{G. Sala}%
\thanks{These authors contributed equally}
\affiliation{Spallation Neutron Source, Second Target Station, Oak Ridge National Laboratory, Oak Ridge, Tennessee 37831, USA}

\author{M. B. Stone}
\affiliation{Neutron Scattering Division, Oak Ridge National Laboratory, Oak Ridge, Tennessee 37831, USA}

\author{V. O. Garlea}
\affiliation{Neutron Scattering Division, Oak Ridge National Laboratory, Oak Ridge, Tennessee 37831, USA}

\author{S. Calder}
\affiliation{Neutron Scattering Division, Oak Ridge National Laboratory, Oak Ridge, Tennessee 37831, USA}

\author{Jie Chen}
\affiliation{International Center for Materials Nanoarchitectonics (WPI-MANA), National Institute for Materials Science, 1-1 Namiki, Tsukuba, Ibaraki 305-0044, Japan}
\affiliation{Graduate School of Chemical Sciences and Engineering, Hokkaido University, North 10 West 8, Kita-ku, Sapporo, Hokkaido 060-0810, Japan
}

\author{Hiroyuki K. Yoshida}
\affiliation{Department of Physics, Faculty of Science, Hokkaido University, Sapporo, Hokkaido 060-0810, Japan}
\affiliation{International Center for Materials Nanoarchitectonics (WPI-MANA), National Institute for Materials Science, 1-1 Namiki, Tsukuba, Ibaraki 305-0044, Japan}

\author{Shuhei Fukuoka}
\affiliation{Department of Physics, Faculty of Science, Hokkaido University, Sapporo, Hokkaido 060-0810, Japan}

\author{Jiaqiang Yan}
\affiliation{Materials Science and Technology Division, Oak Ridge National Laboratory, Oak Ridge, Tennessee 37831, USA}

\author{Clarina dela Cruz}
\affiliation{Neutron Scattering Division, Oak Ridge National Laboratory, Oak Ridge, Tennessee 37831, USA}

\author{Mao-Hua Du}
\affiliation{Materials Science and Technology Division, Oak Ridge National Laboratory, Oak Ridge, Tennessee 37831, USA}

\author{David S. Parker}
\affiliation{Materials Science and Technology Division, Oak Ridge National Laboratory, Oak Ridge, Tennessee 37831, USA}

\author{Hao Zhang}
\affiliation{Materials Science and Technology Division, Oak Ridge National Laboratory, Oak Ridge, Tennessee 37831, USA}%
\affiliation{Department of Physics and Astronomy, University of Tennessee, Knoxville, TN 37996}

\author{C. Batista}
\affiliation{Neutron Scattering Division, Oak Ridge National Laboratory, Oak Ridge, Tennessee 37831, USA}
\affiliation{Department of Physics and Astronomy, University of Tennessee, Knoxville, TN 37996}

\author{Kazunari Yamaura}
\affiliation{International Center for Materials Nanoarchitectonics (WPI-MANA), National Institute for Materials Science, 1-1 Namiki, Tsukuba, Ibaraki 305-0044, Japan}
\affiliation{Graduate School of Chemical Sciences and Engineering, Hokkaido University, North 10 West 8, Kita-ku, Sapporo, Hokkaido 060-0810, Japan
}%

\author{A. D. Christianson}
\email{christiansad@ornl.gov}
\affiliation{Materials Science and Technology Division, Oak Ridge National Laboratory, Oak Ridge, Tennessee 37831, USA}%

\date{\today}

\begin{abstract}
We investigate the physical properties of the Shastry-Sutherland lattice material \bnzo{}. Neutron diffraction, magnetic susceptibility, and specific heat measurements reveal antiferromagnetic order below 1.65 K. The magnetic order is found to be a 2-$\boldsymbol{Q}$ magnetic structure with the magnetic moments lying in the Shastry-Sutherland lattice planes comprising the tetragonal crystal structure of \bnzo. The ordered moment for this structure is 1.9(1) $\mu_B$ per Nd ion.  Inelastic neutron scattering measurements reveal that the crystal field ground state doublet is well separated from the first excited state at 8 meV. The crystal field Hamiltonian is determined through simultaneous refinement of models with both the LS coupling and intermediate coupling approximations to the inelastic neutron scattering and magnetic susceptibility data.  The ground state doublet indicates that the magnetic moments lie primarily in the basal plane with magnitude consistent with the size of the determined ordered moment.  

\end{abstract}

\maketitle

\section{\label{intro}Introduction}

The Shastry-Sutherland model\cite{shastry_original,Miyahara_2003} is one of the quintessential models of two dimensional quantum magnetism.  This model is a spin-1/2 square lattice model with a twist--not all of the squares have a next nearest exchange interaction.  Rather, only every other square has a next neighbor interaction.  This turns out to be an important modification which generates novel physical behavior, but still allows for a comprehensive understanding for rather broad values of the two exchange interactions in the model\cite{Miyahara_2003}.  Recently, the Shastry-Sutherland model has gained renewed attention due to the possibility of topological character of the magnetic excitations\cite{scbo_theory,scbo_triplon}.

\begin{figure}
\includegraphics[width=0.97\columnwidth]
                {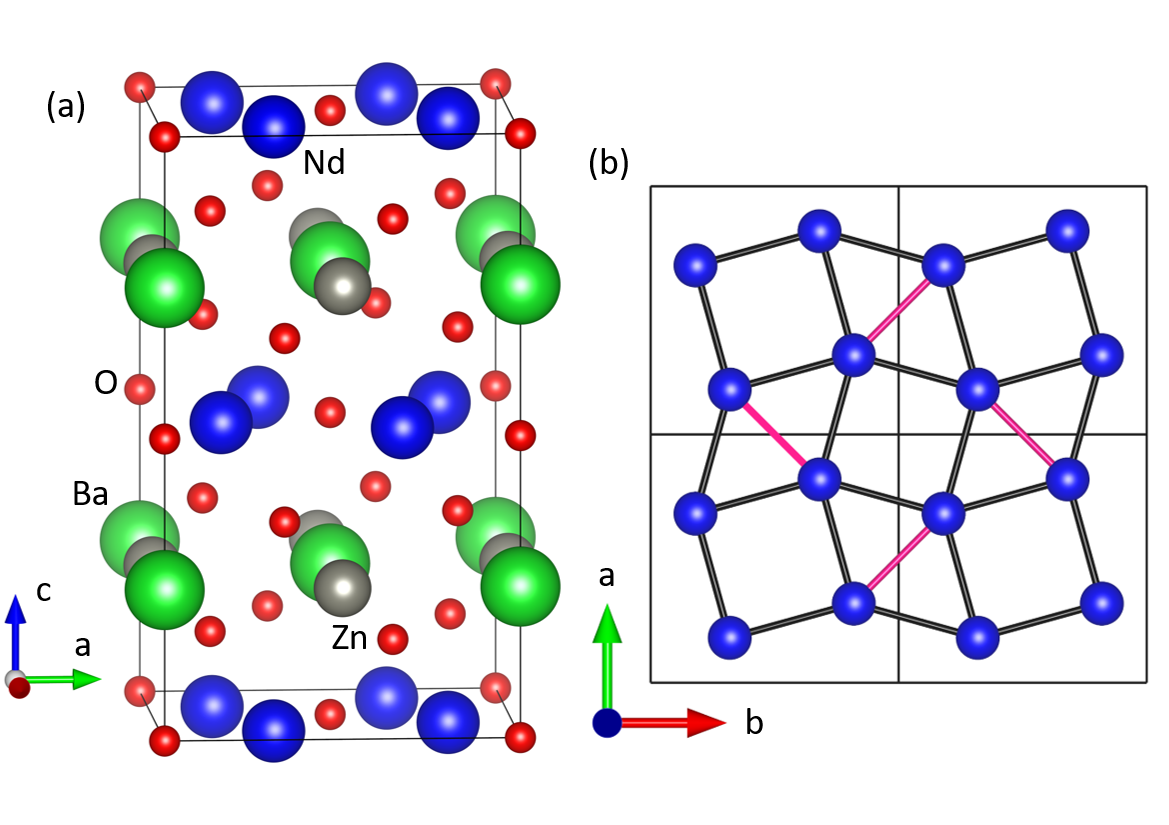}
\caption{
\label{structure}
(a) The crystal structure of \bnzo{}.  One chemical unit cell is depicted.   \bnzo{} crystallizes in the tetragonal space group $I$4/$mcm$ with lattice parameters at 300 K of $a$ = 6.763 \AA \ and $c$ = 11.542 \AA, respectively.  (b) The Shastry-Sutherland lattice formed by a plane of Nd$^{3+}$ ions in \bnzo{}.  The nearest neighbor Nd-Nd distance is 3.46 \AA, pink bonds.   The next nearest neighbor distance is 3.51 \AA{}, black bonds.  The shortest Nd-Nd distance along the c-axis is 5.92~\AA.}
\end{figure}

Despite the interesting physics displayed by the Shastry-Sutherland model, magnetic materials which can be mapped onto the model are rare.  The most notable realization is SrCu$_2$(BO$_3$)$_2$\cite{smith_1991,kageyama_1999,Miyahara_2003} which hosts a lattice with dimers or strongly coupled spin pairs of Cu$^{2+}$ ions arranged orthogonally to one another as illustrated schematically in Fig.~\ref{structure}(b). This lattice geometry is topologically equivalent to the Shastry-Sutherland lattice where the nearest neighbor and next nearest neighbor interactions are reversed so that the intra-dimer and inter-dimer interactions are denoted by $J$ and $J'$ respectively. The ground state of SrCu$_2$(BO$_3$)$_2$ is a dimer singlet state \cite{kageyama_1999,Miyahara_2003}. However, the ratio of the two magnetic interactions $\delta = J'/J \sim 0.63$ is near the quantum critical point between the dimer singlet and the plaquette singlet states\cite{matsuda_2013,Miyahara_2003}. Indeed, several studies have shown how the ground state can be modified by tuning $\delta$ though the application of pressure\cite{haravifard_2014,sakurai_2018}. Recently, theoretical modeling\cite{scbo_theory,scbo_triplon} and inelastic neutron scattering studies\cite{scbo_triplon} have demonstrated multiple triplon bands and possible topological phase transitions between phases with different topologically nontrivial bands.

Beyond SrCu$_2$(BO$_3$)$_2$, there are few examples of magnetic Shastry-Sutherland lattice materials which have been investigated in detail.  Two families of insulating materials where investigations have been made are:  The RB$_4$ (R=rare earth) family of materials\cite{berrada_1976,siemensmeyer_2008,ye_2017} and a large family of materials with generalized chemical formula of the form BaR$_2$TO$_5$ (R=rare earth, T=transition metal) \cite{kaduk_1999,taniguchi_2005,sala2017,ishii_2020}.  There have also been investigations of metallic Shastry-Sutherland lattice materials such as Yb$_2$Pt$_2$Pb, where the spin Hamiltonian governing the magnetic behavior is quasi-one-dimensional\cite{wu_2016} and materials in the family R$_2$T$_2$In (R=rare earth, T=transition metal)\cite{zaremba_2004,tobash_2005,fischer_2000}.

Here we focus on the insulating material \bnzo{}, a member of the BaR$_2$TO$_5$ family (the chemical formula for the family is sometimes given as R$_2$BaTO$_5$ in the literature), as a model magnetic system for study. \bnzo{} hosts Shastry-Sutherland layers of Nd$^{3+}$ ions\cite{TAIBI1990233}. The crystal structure has tetragonal symmetry and crystallizes in space group $I4/mcm$. The structure can be understood as consisting of  Shastry-Sutherland layers of Nd$^{3+}$ ions alternating with  nonmagnetic Ba-Zn spacing layers as shown in Fig.~\ref{structure}(a). The Nd$^{3+}$ - Nd$^{3+}$ distances in the Shastry-Sutherland layer are 3.46~\AA{} and 3.51~\AA{} as shown in Fig.~\ref{structure}(b). The shortest Nd-Nd distance along the $c$-axis is 5.92~\AA{} suggesting that the Nd layers are well isolated from each other.   Kageyama $et\ al.$ report a large negative Weiss temperature $\theta_{\rm W}$ = -44.5 K and attribute a peak in the susceptibility at 2.4 K to the onset of antiferromagnetic order\cite{Kageyama2002}. If the effects of crystal field level spitting are neglected, the large ratio $|\theta_{\rm W}$/$T_{\rm N}|$ = 19 suggests strong frustration.  The crystal field level splitting in \bnzo{} has been previously investigated with optical techniques\cite{taibi_cf_1989,Klimin2001,Rudowicz2007} with some differences in the observed crystal field levels and conventions used to determine the crystal field parameters.

\begin{figure}
\includegraphics[width=0.90\columnwidth]
                {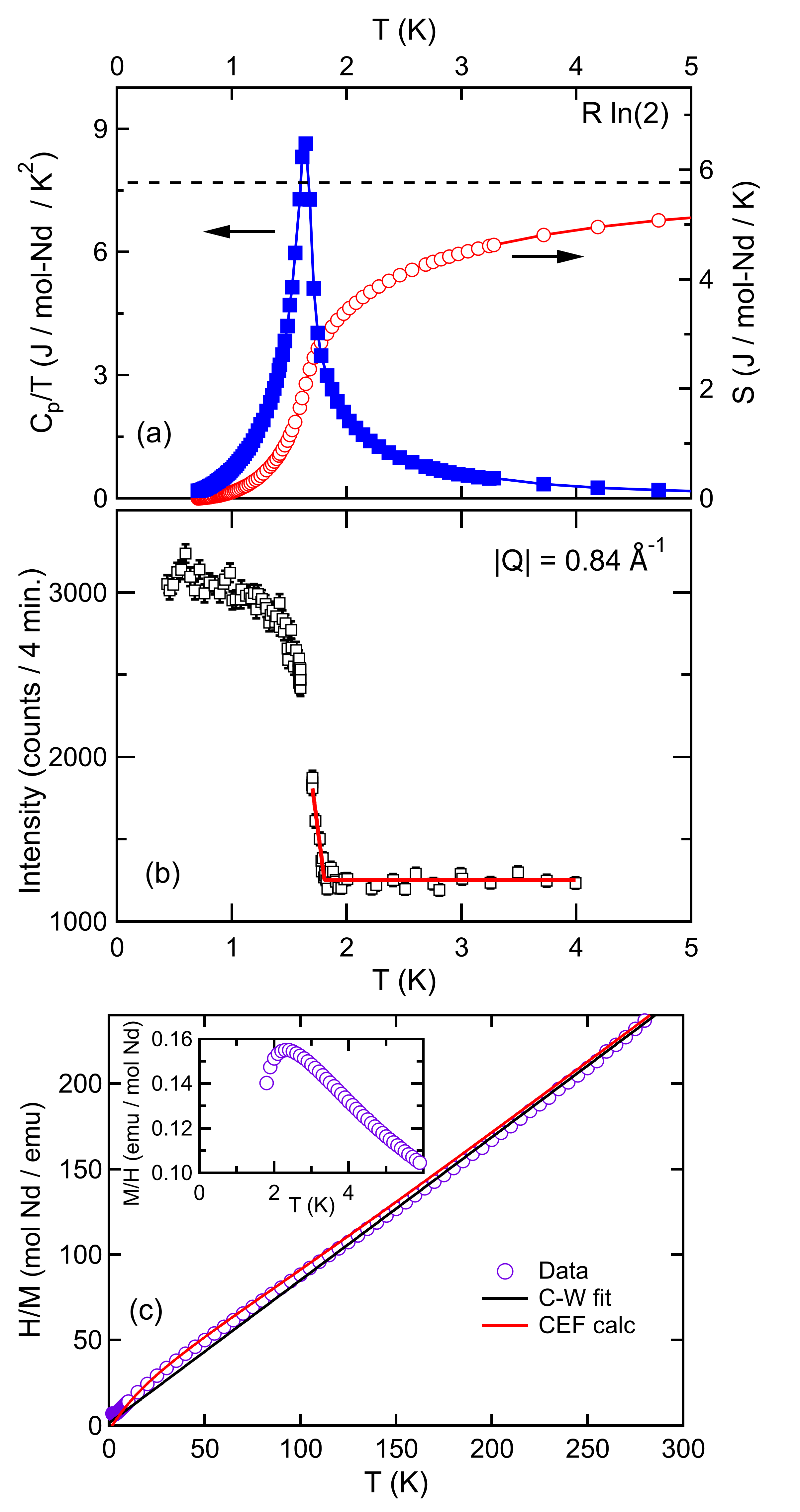}
\caption{
\label{bulk}
(a) Specific heat divided by temperature ($C_p/T$) and entropy ($S$) of \bnzo{}. (b)Neutron diffraction data showing the temperature dependence of the powder averaged peak intensity of the (0.5 0.5 1) magnetic peak and symmetry equivalents. The red line is a power law fit which yields $T_N$=1.80(5) K. (c) The inverse magnetic susceptibility ($H/M$) with an applied field of 0.1 T.  The inset shows the low temperature behavior of the magnetic susceptibility ($M/H$). The black solid line is the result of a Curie-Weiss fit for $100 \leq T \leq 300$~K as described in the text. The red solid line is the calculated susceptibility based upon the results of the crystal field analysis.}
\end{figure}

Given the potential for topological behavior of the Shastry-Sutherland lattice layers in the BaR$_2$TO$_5$ family a comprehensive characterization of the physical properties of representative members is important.  To that end,  in this paper we study \bnzo{} with a combination of neutron diffraction, inelastic neutron scattering, magnetization, and specific heat capacity measurements.  Specific heat measurements show that a transition to long range order occurs at 1.65 K.  Using neutron diffraction data we find the magnetic structure is a 2-$\boldsymbol{Q}$ antiferromagnetic  structure with an ordered moment of 1.9(1) $\mu_B$ per Nd ion.  Using inelastic neutron scattering we measure the crystal field excitations and determine the crystal field Hamiltonian.   The crystal field Hamiltonian explains the magnetic susceptibility data indicating that the rather large frustration ratio of 19 is at least in part the result of crystal field level spitting and further indicates that the moments on the Nd ions are preferentially orientated perpendicular to the vector connecting Nd nearest neighbors (pink bond in Fig. \ref{structure}(b)) in agreement with the magnetic structure.

\begin{figure}
\includegraphics[width=0.99\columnwidth]
                {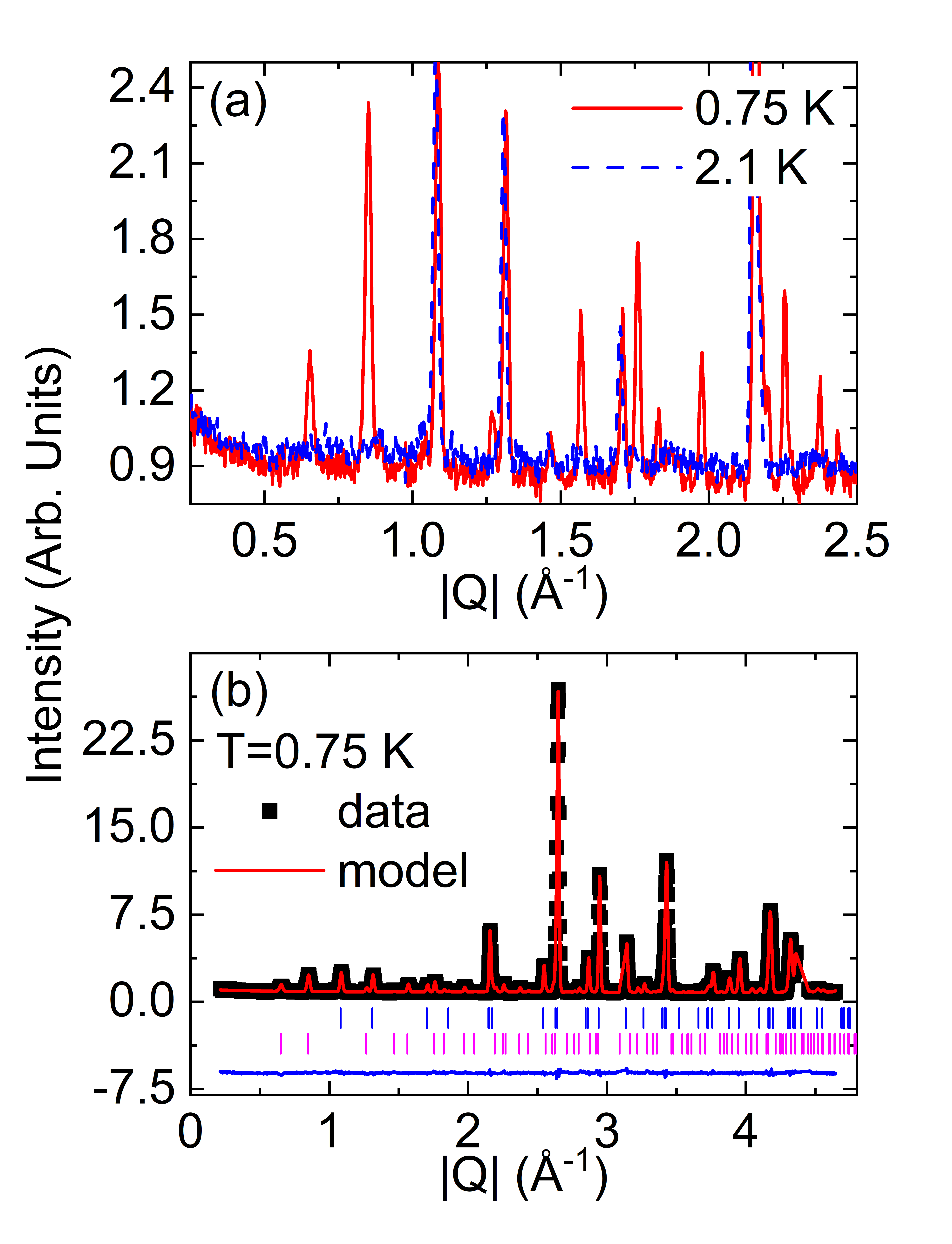}
\caption{
\label{magdiffraction}
(a) Neutron diffraction data in the low Q region above (T=2.1 K) and below (T=0.75 K) the magnetic phase transition.  (b) Fit of the magnetic and nuclear structure models to the data as described in the text. The top series of blue tick marks indicates structural reflections and the bottom series of pink tick marks indicates magnetic reflections.  The difference between the data and model is displayed by the blue line at the bottom of the panel. Regions of data where Al peaks occur (arising due to scattering from the sample holder) have been excluded from the plots above and from the refinements of the structural models.}
\end{figure}

\section{Experimental Details}

Samples of \bnzo{} and \lnzo{} were synthesized by pelletizing a finely ground mixture of stoichimetric amounts of BaCO$_3$, Nd$_2$O$_3$, and ZnO. The pellet was heated at 850$^{\circ}$C for 2 days, 950$^{\circ}$C for 3 days and 1000$^{\circ}$C for 1.5 days in air. After each firing, the sample was removed from the furnace and reground and pelletized\cite{kaduk_1999}.

Heat capacity measurements were performed on a 210~$\mu$g pellet of \bnzo{} using the thermal relaxation method with a homemade calorimetry system equipped with a $^3$He cryostat\cite{Fukuoka2013}. The pellet was fixed onto the sample stage using a small amount of Apiezon N grease.   Magnetic susceptibility measurements were performed using a SQUID magnetometer (Magnetic Property Measurement System, Quantum Design) in the temperature range 2 to 300 K and in an applied magnetic field of 0.1 T.

Neutron powder diffraction measurements to study the crystal and magnetic structures of \bnzo{} were performed using the HB-2A powder diffractometer at the High Flux Isotope Reactor (HFIR).  Additional characterization at 300 K was performed using the POWGEN diffractometer at the Spallation Neutron Source (SNS) at Oak Ridge National Laboratory and synchrotron x-ray diffraction with the BL15XU beamline at SPring-8 in Japan (additional details are provided in Appendix \ref{struct_prop}).  For the HB-2A measurements the sample was loaded into a Al sample can with He exchange gas. HB-2A measurements with $\lambda$ = 2.41 \AA~were conducted for temperatures between 0.5 and 3 K using a cryostat with a $^{3}$He insert.  To extract an accurate temperature dependence (Fig. \ref{bulk}(b)), an additional neutron diffraction experiment was conducted with HB2A using a pressed pellet of the same sample used in the other neutron scattering studies.   For technical reasons, the temperature dependence was conducted at zero field inside of a superconducting magnet with a $^{3}$He insert, which resulted in larger background scattering than for the data used to refine the magnetic structure (Fig. \ref{magdiffraction}). The magnetic structure refinement was performed with the software FullProf~\cite{Fprof}.  The figures depicting the crystal structure and magnetic structure were produced with VESTA\cite{vesta}

Inelastic neutron scattering measurements were performed using the SEQUOIA spectrometer at the SNS~\cite{seq}. \bnzo{}, \lnzo{} and an empty Al sample can were loaded in a three sample changer located on a bottom loading CCR. \lnzo{} and the empty can were used to
subtract the phonon contribution from the measured scattering intensity of the \bnzo{} sample. Measurements were acquired with neutron incident energies, E$_i$s, of 45, 60, 150 and 500 meV.  The E$_i$=45 meV data were collected in high resolution mode with the remainder of the data collected in the high flux mode of the instrument over the temperature range $5 \leq T \leq 200$ K. The highest incident energy was used to investigate the location of higher J-multiplets which provides an additional constraint on the crystal field refinement.

\begin{figure}
\includegraphics[width=0.96\columnwidth]
                {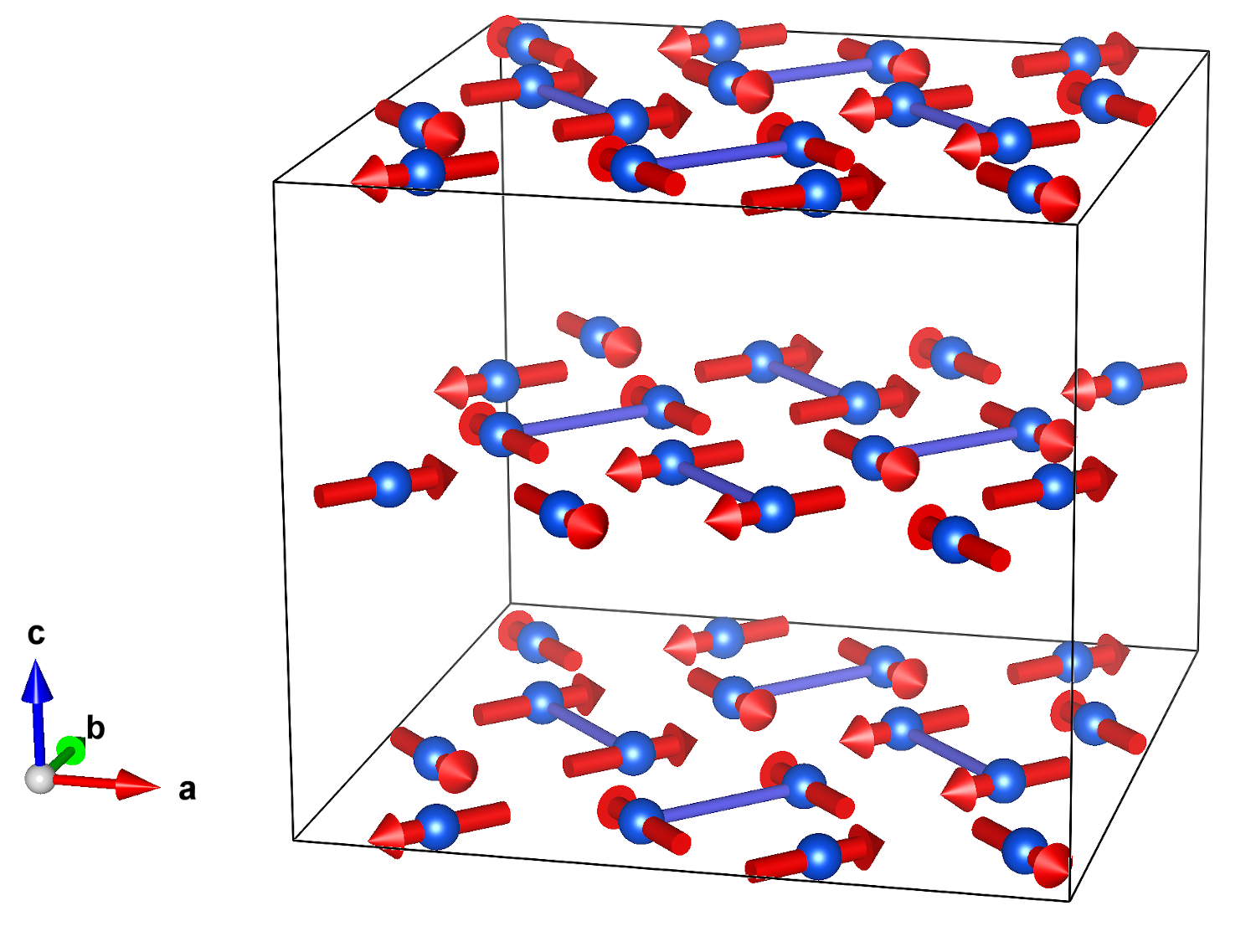}
\caption{
\label{magstruct}
The magnetic structure of \bnzo{}. The structure is described by the magnetic space group $P_C4/nnc$ ($\#126.385$) with lattice parameters of 2a, 2b, and c referenced to the lattice parameters of paramagnetic space group. The two wave-vectors defining this structure, (0.5,0.5,0) and (0.5,-0.5,0), operate on different nearest neighbor dimers to form an orthogonal spin configuration of ferromagnetic dimmers.}
\end{figure}

\section{Thermodynamic Measurements}
Figure \ref{bulk}(a) displays the temperature dependent heat capacity divided by temperature, $C_p/T$ and entropy, $S$. The entropy was estimated by integrating the $C_{\rm p}/T$ vs $T$ curve from 0.7 K to 10 K. Because the contribution of the lattice heat capacity is much smaller than that of the magnetic heat capacity below 10 K, the phonon contribution is neglected. The heat capacity data indicates a phase transition at $T_N$ = 1.65 K (Fig. \ref{bulk}(a)). The temperature dependence of the entropy shows that a large portion of the entropy is released well above the transition temperature and integration up to 10 K is required to reach the value of Rln2.  This indicates that the ground state is a doublet which is expected for a Kramers ion such as Nd$^{3+}$.  We will return to this point later with a detailed discussion of the crystal field splitting in \bnzo{}.  Additionally, the entropy reveals that the interactions generating collective behaviour are significant well above the long range ordering temperature of 1.65 K suggesting a modest degree of frustration and/or low dimensional behavior.

\begin{figure*}
\includegraphics[width=1.95\columnwidth]
                {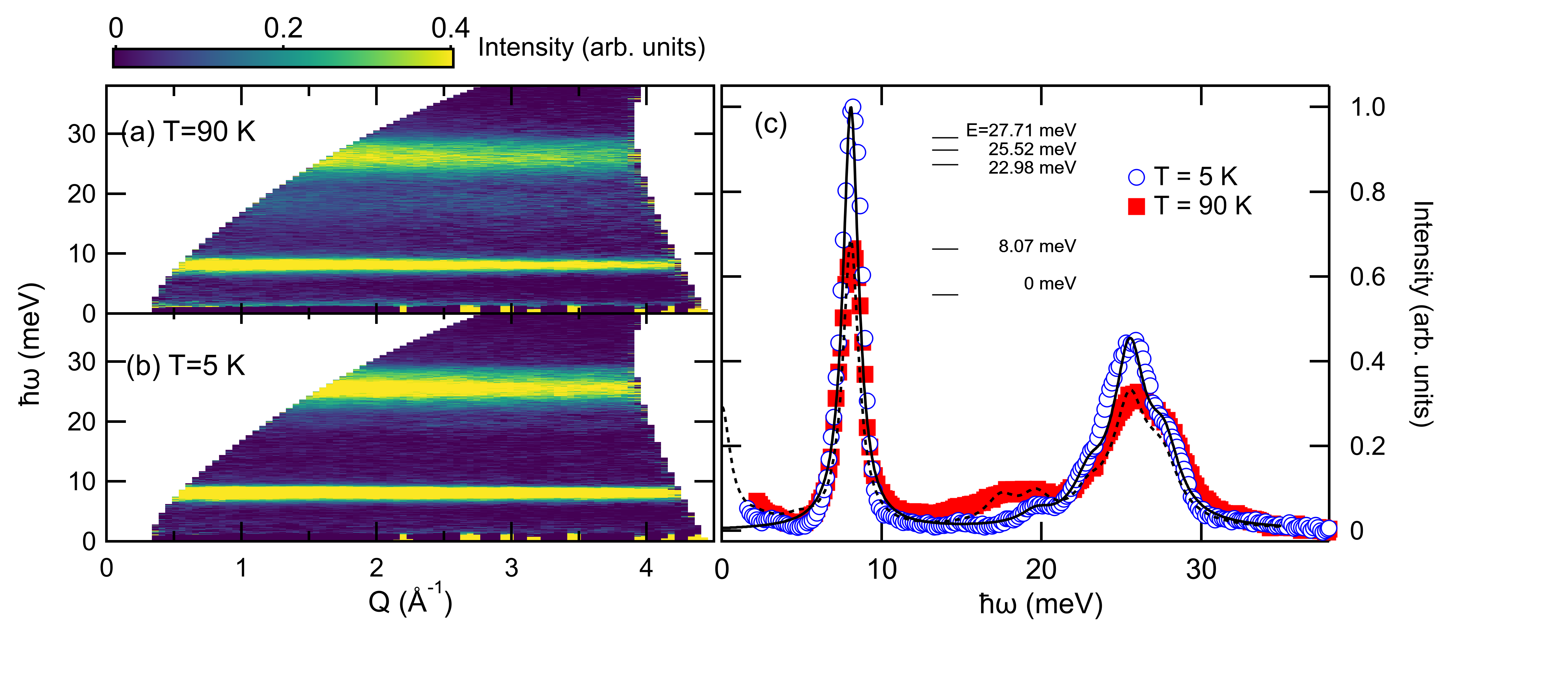}
\caption{
\label{crystalfield}
Crystal field excitations in \bnzo{}. (a) and (b) Intensity maps of the inelastic neutron data collected with SEQUOIA using $E_i=45$ meV at T=90 K (a) and T= 5 K (b). The data show a sharp crystal field excitation at $\hbar\omega=8$ meV, and three
excitations in the range $21\leq \hbar\omega \leq 28$ meV. The intensity decrease at T=90 K is due to the depopulation of the GS state, in favor of excited states. A transition between the first excited state and one of the higher levels is visible near 20 meV in (a). (c) Cuts along $\hbar\omega$  with $Q$ integration range $0.5 \leq |Q| \leq 3.0$ \AA$^{-1}$ compared to the final crystal field model.  The nonmagnetic background has been removed from the data displayed here by subtracting properly normalized data collected under the same conditions of the nonmagnetic analog \lnzo{}. The refined energy levels for the $T=5$~K measurement based upon the LS coupling scheme are illustrated in the inset of (c).
}
\end{figure*}

The temperature dependence of the magnetic susceptibility measurements is shown in Fig. \ref{bulk}(c). The inset shows the low temperature behavior of the susceptibility indicating a broad maximum at 2.3 K which is close to the value of 2.4 K reported by \cite{Kageyama2002}.  This temperature corresponds to the onset of a gradual increase in the heat capacity suggesting that the antiferromagnetic correlations grow from 2.4 K and that long range magnetic order occurs at a temperature corresponding to the peak in the heat capacity.  The neutron scattering data shown in Fig 2(b), and discussed further below, supports this.  The temperature dependent $Q=0.84$~\AA$^{-1}$ magnetic Bragg peak scattering intensity in Fig. 2(b) is fit well by a power law assuming the mean field value, $\beta=0.5$, for the critical exponent and yielding a transition temperature of $T_N=1.80(5)$~K. The small difference between the values of the transition temperature derived from the specific heat and neutron scattering measurements may be related to incomplete thermalization of the large sample used for the neutron diffraction measurements and/or a reflection of short range correlations existing above the ordering temperature.  

\begin{table}
\begin{tabular}{c|cccccccccc} \hline \hline
 Atom & x & y & z & usio\\ 
     \hline
 Ba  & 0 & 0 & 0.25 & 0.5(1) \\ 
 Nd1 & 0.1736(2) & 0.6736(2) & 0 & 0.11(6) \\ 
 Zn & 0 & 0.5 & 0.25 & 0.33(9) \\ 
 O1 & 0 & 0 & 0 & 0.3(1) \\ 
 O2 & 0.3547(1)	& 0.8547(1) & 0.1312(1) & 0.66(6) \\ 
     \hline 
\end{tabular}   

\vspace*{5mm}
\begin{tabular}{c|cccccccccc} \hline \hline
  Lattice parameters & $a_m$,$b_m$=13.4822(2)~\AA & $c_m$=11.5296(2)~\AA  \\
  Magnetic moment & $\mu_{Nd}$ = 1.9(1)$\mu_B$ \\
  Magnetic R-factor &  19  \\
  R$_{wp}$ & 7.4 (2.1 K) & 7.7 (0.75 K) \\
 \hline \hline
\end{tabular}
\caption{
\label{tab: magrefine}
Low temperature refinement parameters. For the refinement of the magnetic structure atom positions and thermal factors where held fixed to the values determined from refining the structural model at 2.1 K and are given in the table.  The refinement parameters pertaining to the magnetic structure are determined from refinements to data collected at 0.75 K. Additional information about the symmetry of magnetic structure is provided in Appendix \ref{ap_mag_sg}.  }
\end{table}

The inverse susceptibility shows almost linear temperature dependence between 100 and 300 K.  Fits to a Curie-Weiss law with and without a temperature independent term results in a range of Curie constants and Weiss temperatures which vary significantly depending on the range of data selected for the fit.  As explained in more depth in the section on crystal field splitting, a proper description of the susceptibility requires a model which considers the thermal population of the crystal field levels.  For completeness and to give a point of reference for future studies, the Curie constant and Weiss temperature determined by fitting the susceptibility for 100 $\leq$ T $\leq$ 300 K are C= 1.24 emu/mol Nd and $\Theta_{CW}$=-7.9 K.

\section{Magnetic Structure}

To determine the nature of the magnetic transition at $T_N$, neutron diffraction measurements were performed using the HB2A powder diffractometer.  The neutron diffraction data are shown in Figs. \ref{bulk}(b) and \ref{magdiffraction}.  Below $T_N$ these measurements reveal an additional set of reflections arising due to magnetic order of the Nd magnetic moments (Fig. \ref{magdiffraction}(a)).  The additional set of peaks can be indexed by wave vectors of the type (0.5 0.5 0).  Analysis to determine symmetry allowed magnetic structures was performed with the Sarah\cite{sarah} and ISODISTORT\cite{cambell_2006} software packages. To determine the magnetic space group of the structure we employed the k-SUBGROUPSMAG program at the Bilbao Crystallographic Server\cite{bilbao}. With a propagation vector of (0.5 0.5 0), symmetry requires that the single Nd site of the parent structure split into two orbits (or magnetically non-equivalent sites, which correspond to different set of dimers). Initial refinements of single-$\boldsymbol{Q}$ magnetic structures resulted in spin configurations with large moments on one Nd orbit whereas the other had essentially a zero moment (Magnetic space group $C_{A}mca$).  Solutions of this type were rejected because no crystallographic distortion was detected in the neutron diffraction data indicating that the two Nd sites remain in identical chemical environments.   As discussed below, the Nd$^{3+}$ ions have a substantial local moment of 2.17 $\mu_B$ from the ground state crystal field doublet which should result in a strong diffuse scattering signal in the case where 1/2 of the magnetic moments do not undergo long range magnetic order.  Diffuse scattering corresponding to this case was not observed in the neutron diffraction pattern. 

With the aforementioned considerations in mind a 2-$\boldsymbol{Q}$ magnetic structural model was fit to the data.  This magnetic structure is characterized by the two propagation vectors (0.5 0.5 0) and (0.5 -0.5 0) acting on different dimer pairs to produce an orthogonal spin configuration as shown in Fig. \ref{magstruct}. Table \ref{tab: magrefine} summarizes the structural and magnetic refinement results at low temperature.   The moments on the Nd$^{3+}$ ions are found to point along [110] and symmetry equivalent directions with a magnitude of 1.9(1) $\mu_B$ and are always perpendicular to the vector connecting nearest neighbor Nd ions. The magnetic space group for this structure is $P_{C}4/nnc$ with lattice parameters of $a_m$,$b_m$=2$a$ and $c_m$ = $c$, where $a$ and $c$ are the lattice parameters of the paramagnetic structure. Note that the double-$\boldsymbol{Q}$ structure recovers the symmetry of the parent structure and all magnetic ions are described with a single Wyckoff position.  We also used simulated annealing as a second approach for determining the magnetic structure. The analysis was performed on the integrated magnetic intensities obtained after subtracting the nuclear contribution measured in the paramagnetic state. The paramagnetic scattering that follows the Nd$^{3+}$ magnetic form factor decay with momentum transfer Q was property accounted for in the subtraction. We defined a magnetic unit cell as $2a$,$2b$,$c$ and allowed the magnetic moment to vary freely inside the $ab$ plane.  This method led to the same spin configuration of ferromagnetic dimers with magnetic moments oriented perpendicular to the bond direction.  The value and orientation of the ordered moment compares well with that expected from the ground state crystal field doublet as will be discussed further in Sec. \ref{cfsec}.  This magnetic structure has ferromagnetic dimers rather than the dimer singlets of the prototypical Shastry-Sutherland model material \scbo{}\cite{kageyama_1999,Miyahara_2003}.

\section{Crystal Field Level Splitting\label{cfsec}}

We examine the local properties of the Nd$^{3+}$ ions in \bnzo{} by studying the crystal field excitation spectrum to determine the crystal field Hamiltonian. We analyze the inelastic neutron scattering data using the formalism described by Wybourne~\cite{Wybourne} based on the Racah tensor operator method~\cite{Racah1,Racah2,Racah3,Racah4}. The Nd ions occupy atomic positions with $C_{2v}$ point symmetry. The coordinate system was rotated by $45^{\circ}$ around the $\hat{z}$-axis (crystallographic c-axis) to achieve Prather's convention~\cite{Prather}, which ensures the lowest number of crystal field parameters to approximate the Coulomb potential, thus we can write the crystal field Hamiltonian as:
\begin{eqnarray}
H = B_2^0\hat{C}_2^0 + B_2^2(\hat{C}_2^2+\hat{C}_2^{-2}) + B_4^0\hat{C}_4^0 + B_4^2(\hat{C}_4^2+\hat{C}_4^{-2}) +
\nonumber
\\
B_4^4(\hat{C}_4^4-\hat{C}_4^{-4}) + B_6^0\hat{C}_6^0 + B_6^2(\hat{C}_6^2+\hat{C}_6^{-2}) + 
\nonumber
\\
B_6^4(\hat{C}_6^4-\hat{C}_6^{-4}) + B_6^6(\hat{C}_6^6+\hat{C}_6^{-6}) \qquad
\label{eq: 1}
\end{eqnarray}
where $B_n^m$ and $C_n^m$ are Wybourne's crystal field parameters and relative operators respectively. The intermediate
coupling formalism has been employed in our analysis to verify the possible overlap of the ground state (GS) eigenfunctions with higher J-multiplets. Once $H$ is diagonalized, the crystal field intensities can be calculated as shown in Ref.~\cite{sala_ndo}. Given that the total angular momentum of Nd$^{3+}$ is $J=9/2$ we expect a total of $5$ doublets, hence $4$ crystal field transitions within the GS multiplet.

\begin{table} 
\begin{tabular}{c|cc}
\hline\hline
$B_n^m$ & LS-coupling (meV) & Intermediate (meV)\\
\hline
$B_2^0$ & $-55(1)$ \quad & $-61(1)$ \\
$B_2^2$ & $-47.0(3)$ \quad & $-50.7(3)$ \\
$B_4^0$ & $1.03(6)$ \quad & $1.03(5)$\\
$B_4^2$ & $-13(1)$ \quad & $-14(1)$ \\
$B_4^4$ & $-75.5(3)$ \quad & $-75.8(3)$ \\
$B_6^0$ & $1.11(8)$ \quad & $1.07(7)$ \\
$B_6^2$ & $16.6(5)$ \quad & $19.2(6)$ \\
$B_6^4$ & $-4.1(5)$ \quad & $-4.4(5)$ \\
$B_6^6$ & $48(3)$ \quad & $49(3)$ \\
\hline\hline
\end{tabular}
\caption{
\label{tab: 1}
Tabulated crystal field parameters in units of meV determined for the LS-coupling and Intermediate coupling approximations.}
\end{table}

\begin{table*} 
\begin{tabular}{c|cccccccccc}
\hline\hline
$m_J$ & $0.0$ & $0.0$ & $8.07$ & $8.07$ & $22.98$ & $22.98$ & $25.52$ & $25.52$ & $27.71$ & $27.71$\\
\hline
$-9/2$ & 0.0987 & & & 0.395 & & 0.166 & & -0.702 & & 0.559 \\
$-7/2$ & & -0.527 & 0.0669 & & -0.167 & & 0.516 & & -0.651 & \\
$-5/2$ & -0.596 & & & 0.0475 & & 0.617 & & 0.370 & & 0.353 \\
$-3/2$ & & 0.486 & 0.565 & & -0.512 & & -0.211 & & -0.372 & \\
$-1/2$ & 0.347 & & & -0.720 & & 0.549 & & -0.244 & & -0.0214 \\
$1/2$ & & 0.347 & -0.720 & & -0.549 & & 0.244 & & -0.0214 & \\
$3/2$ & 0.486 & & & 0.565 & & 0.512 & & 0.211 & & -0.372 \\
$5/2$ & & -0.596 & 0.0475 & & -0.617 & & -0.370 & & 0.353 & \\
$7/2$ & -0.527 & & & 0.0669 & & 0.167 & & -0.516 & & -0.651 \\
$9/2$ & & 0.0987 & 0.395 & & -0.166 & & 0.702 & & 0.559 & \\
\hline\hline
\end{tabular}
\caption{
\label{tab: 2}
Tabulated wave functions of the crystal field states in \bnzo~obtained within the LS-coupling approximation. The crystal field energies (in meV) are tabulated
horizontally, the $m_J$-values of the ground-state multiplet vertically. Only coefficients of the wave functions $> 10^{-3}$ are shown. }
\end{table*}

 The inelastic neutron scattering measurements performed with SEQUOIA are displayed in Fig. \ref{crystalfield}.  The measurements  at T=5 and 90 K identified a series of crystal field excitations in the range $8\leq \hbar\omega \leq 30$ meV, as shown in Fig.~\ref{crystalfield}. Among these excitations, the first excited state is well isolated at $\hbar\omega=8$ meV, while the remaining three are peaked in a range $23\leq \hbar\omega \leq 27$ meV. The resolution of the SEQUOIA spectrometer and the temperature dependence of transitions between excited states allowed the positions of the crystal field levels to be identified (Fig. \ref{crystalfield}(c) and Tab. \ref{tab: 2}). 
 
 Using the procedure described in Ref. \cite{sala_ndo} the inelastic neutron scattering data collected at T=$5$ and T=$90$ K, and the magnetic susceptibility data from $270$ K to $150$ K were simultaneously refined to determine the crystal field parameters of Eq.~\ref{eq: 1}. The calculation was first performed in LS-coupling, then extended to intermediate coupling and verified with the software SPECTRE~\cite{spectre}. The refined crystal field parameters within the LS and intermediate couplings are shown in Tab.~\ref{tab: 1}. There is good agreement between the data and calculation. The overall spectrum is captured by our model as are the transitions between excited states at T$=70$ K. Note that we subtracted the nonmagnetic equivalent of \bnzo, \lnzo, from the data set to eliminate the phonon contribution to the spectrum. Based on our data set we did not find evidence of phonon-crystal field coupling, nevertheless, as will be discussed below, there appear to be several phonon modes which have been previously misidentified as crystal field levels.  

Finally, we report in Tabs.~\ref{tab: 2} and \ref{tab: 3} the wave-functions of the crystal field states for \bnzo, calculated within
the LS and intermediate coupling approximation. Considering that the $J=11/2$ multiplet lays at $\hbar \omega=237.35$ meV, and that the $m_J$ mixing of the GS wave-functions is quite small ($\leq 4\%$), we conclude that the main contributions to the magnetic properties of the sample arise from the GS multiplet with no significant contribution from higher multiplets.

\section{Discussion}

We first discuss the difference between our crystal field model, and the previously reported models for \bnzo~\cite{taibi_cf_1989,Klimin2001}. The previous crystal field measurements obtained by optical data
are significantly different from one another; both authors identified the first excited state of \bnzo~around $\Delta E$=8 meV, consistent with our observations, but they report one or more higher excited states above 30 meV that to not appear to be crystal field levels based on our analysis. Furthermore, they do not benchmark their results against physical properties such as the magnetic susceptibility which as we discuss provides further confirmation of our crystal field model.

To understand the origin of the spectra above 30 meV we checked the Q-dependence of the excitations at $18 \leq \hbar \omega \leq 22$ meV, $33 \leq \hbar \omega \leq 38$ meV and $55 \leq \hbar \omega \leq 59$ meV (see Appendix Fig.~\ref{cfqdep}). A quadratic behavior of these levels as a function of Q is clearly observed, such behavior is the signature of phonon scattering in an inelastic neutron scattering experiment.  To gain additional confidence that the modes are indeed phonons rather than crystal field excitations, DFT calculations were performed (see Appendix \ref{phononcalcs})  Specifically, the levels at $37.815$ meV in~\cite{taibi_cf_1989} and at $35.58$, $57.03$ meV in~\cite{Klimin2001} 
were compared to our DFT calculations for \bnzo. Indeed, we found three phonons at $34.83$ ($A2u$), $37.51$ ($Eu$) and 
$58.84$ ($Eg$) meV that match closely the previous optical measurements (see Appendix Fig.~\ref{dosfigure}).

As an additional check of the crystal field levels reported by Refs. \cite{taibi_cf_1989, Klimin2001}, we investigated if the ground state of these models was consistent with the physical properties of \bnzo{}.  For example, computing the ground state wave-functions in intermediate coupling of \bnzo{} using either Ref.~\cite{taibi_cf_1989}
or Ref.~\cite{Klimin2001} results in a strong linear combination of $0.715|\pm 9/2\rangle + 0.587|\pm5/2\rangle$ and $0.75|\pm 9/2\rangle + 0.514|\pm5/2\rangle$ respectively. This would imply that spins have an Ising-like nature pointing perpendicular to the Shastry-Sutherland lattice, in contradiction with the magnetic structure determined here. 

Based on our analysis all crystal field levels of the ground state (J=9/2) multiplet are found below $30$ meV; with the first excited state at 8.07 meV and the remaining three excited states located in close proximity to each other at $\hbar \omega=22.98$, 
$\hbar \omega=25.52$ and $\hbar \omega=27.71$ meV. Additionally, by computing the crystal field spectrum in intermediate coupling (see Tab.~\ref{tab: 4}), our model reproduces the energies of the first 
J-multiplet $^4I_{11/2}$ at $237 \leq \hbar \omega \leq 260$ meV, and $^4I_{13/2}$ at $480 \leq \hbar \omega \leq 503$ meV in good agreement with what has been measured in Refs.~\cite{taibi_cf_1989,Klimin2001}.

The simultaneous refinement of the energies, integrated intensity and susceptibility imposes strong constraints on the fit, and allowed us to determine a magnetic moment of $\mu=2.17(7)\mu_B$/ion for the ground state doublet, in close agreement with the ordered moment of 1.9 $\mu_B$ of the refined magnetic structure.  The g-factors can also be determined and are g$_z=0.486$, g$_x=1.31$ and g$_y=4.752$ indicating spins laying in the ab plane of the compound. Note that the subscripts indicated on the g-factors refer to the rotated coordinate system used for the crystal field calculation.  These g-factors show that the moments have a strong tendency to point perpendicular to the vector connecting the nearest neighbor Nd ions, 
in agreement with the magnetic structure. Indeed, the 2-$\boldsymbol{Q}$ structure here is likely stabilized by the anisotropy introduced by the crystal field Hamiltonian.

\section{Conclusions}

We have studied the magnetic properties of the Shastry-Sutherland lattice material \bnzo{}.  Long range magnet order occurs below $T_N$= 1.65 K.  The magnetic structure is a 2-$\boldsymbol{Q}$ structure with a magnetic moment of 1.9 $\mu_B$.  The heat capacity and magnetic susceptibility indicate that significant magnetic correlations build above the ordering temperature.  The crystal field Hamiltonian determined here does a good job explaining the physical properties including the magnetic susceptibility above 15 K and the orientation of the magnetic moments. The distribution of magnetic entropy around the transition to long range order indicates the significant potential for collective magnetic excitations in \bnzo{}.  The very low transition temperature and relatively large spin quantum indicate that such excitations would be at energies of the order of 1 meV.  The geometrical frustration and 2-$\boldsymbol{Q}$ structure at play in \bnzo{} will significantly influence these collective excitations.

\begin{acknowledgments}
We thank A. F. May and B. J. Campbell for useful discussions. This work was supported by the U.S. Department of Energy, Office of Science, Basic Energy Sciences, Materials Sciences and Engineering Division. This research used resources at the Spallation Neutron Source and the High Flux Isotope Reactor, Department of Energy (DOE) Office of Science User Facilities operated by Oak Ridge National Laboratory (ORNL). This study was supported in part by JSPS KAKENHI Grants No. JP20H05276, a research grant from Nippon Sheet Glass Foundation for Materials Science and Engineering (Grant No. 40-37), and Innovative Science and Technology Initiative for Security (Grant No. JPJ004596) from Acquisition, Technology \& Logistics Agency (ATLA), Japan.
\end{acknowledgments}

%

%

\clearpage

\appendix

\section{Crystal Field Analysis}

\renewcommand{\thefigure}{A\arabic{figure}}
\renewcommand{\thetable}{A\Roman{table}}
\setcounter{figure}{0}
\setcounter{table}{0}

\begin{figure}[h]
\includegraphics[width=0.95\columnwidth]
                {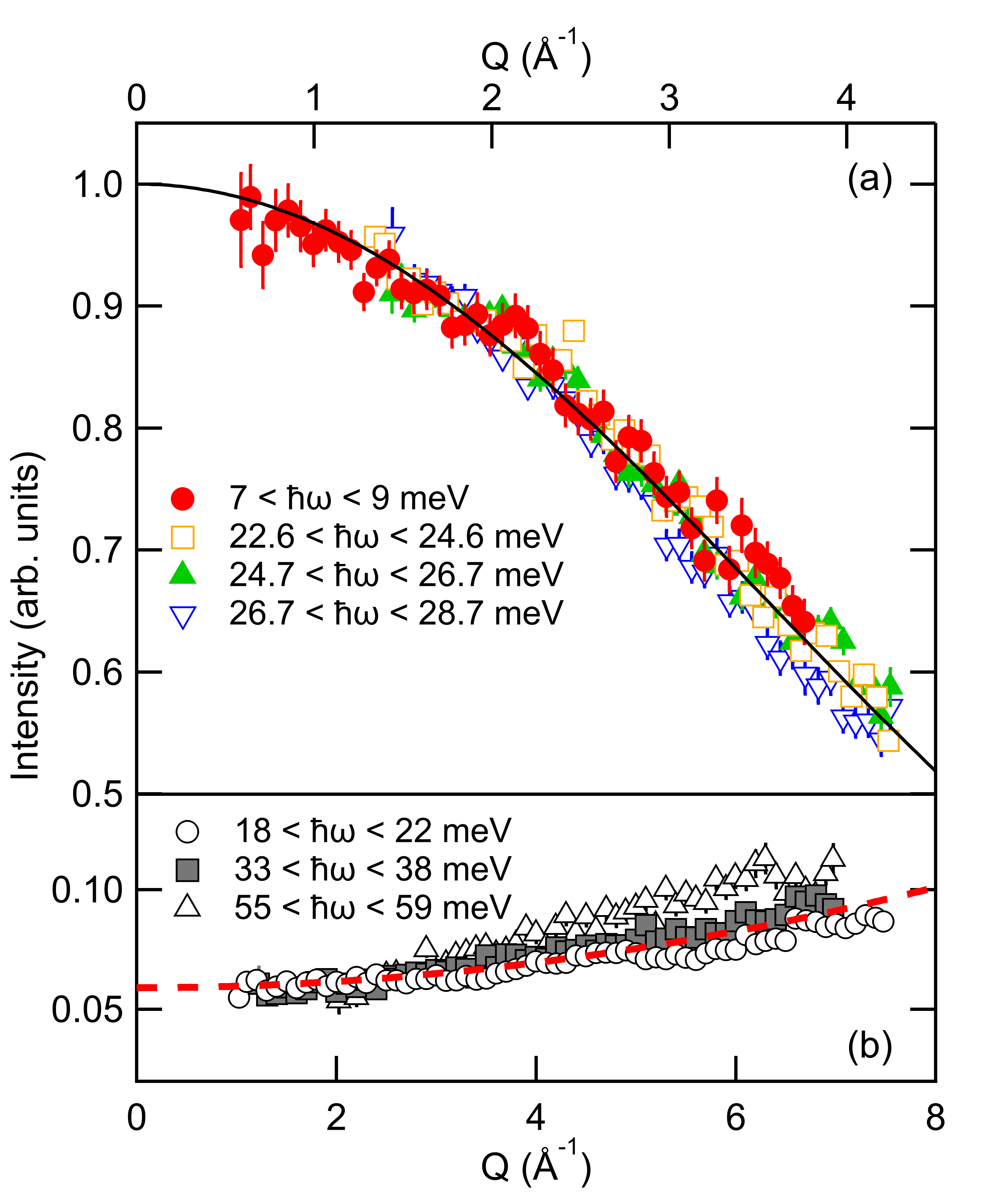}
\caption{
\label{cfqdep}
$Q$-dependence of the scattering intensity of \bnzo{} as measured at $T=5$~K.  (a) The constant energy cuts of the crystal field excitations were scaled by a multiplicative factor to normalize their integrated scattering intensity over their shared range of wave-vector transfer. The good agreement with the fit of all the data (black line), to the square of the Nd$^{3+}$ magnetic form factor, validates their crystal field origin. (b) Constant energy cuts through the inelastic neutron scattering measurements over the range of energy transfer reported in Refs.~\cite{taibi_cf_1989, Klimin2001}. Data have been scaled by a multiplicative pre-factor so that they share a common integrated scattering intensity over a range of measured values of $|Q|$. Dashed red line is a quadratic function with a constant background fit through all these data, highlighting the phonon nature of the excitations.}
\end{figure}

In order to distinguish between crystal field levels and the phonon modes of the \bnzo inelastic neutron scattering measurements, we examined the $|Q|$ dependence of each excitations taking constant energy cuts thought the data set. 
As shown in Fig.~\ref{cfqdep}(a), individual energy cuts centered around $\pm1$ meV on each crystal field level are compared to the square of the magnetic form factor for Nd$^{3+}$ (black line) within an overall multiplicative constant. 
These individual cuts through the measured spectra were scaled to one another with a multiplicative factor so that they would be on the same intensity scale. This comparison highlights the excellent agreement of the $|Q|$ dependence with the magnetic ion form factor, verifying the origin of the scattering.

The same procedure was used to check the origin of the scattering at the energies of the excitations reported in Refs.~\cite{taibi_cf_1989, Klimin2001}. The ranges of energy transfer corresponding to these excitations are shown in Fig.~\ref{cfqdep}(b). These data have also been scaled to appear on the same intensity scale using a multiplicative scale factor. For comparison, we also plot a quadratic function with a constant background as a guide to the eye. The good agreement with this function indicates that the origin of the scattering intensity of these modes is likely due to phonons, and not
crystal field excitations. This result is also independently confirmed by our DFT calculations (Appendix \ref{phononcalcs}).

The crystal field fit was first performed assuming LS-coupling and then extended to intermediate coupling to verify the overlap of the eigenfunctions between the GS J-multiplet and any of the higher J-multiplets. This is an important verification that needs to be done to fully understand the properties of the GS, especially for light rare earth atoms. Both optical measurements showed a huge gap $\geq 200$ meV between the the $^4I_{9/2}$ and the $^4I_{11/2}$ multiplets, thus giving a negligible overlap of the respective eigenfunctions. Indeed, our refinement confirmed an overlap $\leq3$\%. Our calculations have also been independently verified with the software SPECTRE~\cite{spectre}. The tabulated final wave functions of the crystal field levels for \bnzo{} up to the first excited multiplet, are shown in Tab.~\ref{tab: 3}. Finally, a comparison of the calculated energies of the first three J-multiplets with Refs.~\cite{taibi_cf_1989, Klimin2001} is shown in Tab.~\ref{tab: 4}. As shown in the table, our analysis agrees remarkably well with the high energy excitations measured with optical methods.

\begin{table}[h]
\begin{tabular}{c|cccc}
\hline\hline 
\multicolumn{1}{c|}{\begin{tabular}[c]{@{}c@{}} Level \end{tabular}} & \multicolumn{1}{c}{\quad Exp.~\cite{taibi_cf_1989} \quad} & \multicolumn{1}{c}{\quad Exp.~\cite{Klimin2001} \quad} & \multicolumn{1}{c}{\begin{tabular}[c]{@{}c@{}}Exp.\\ (this work)\end{tabular}} & \multicolumn{1}{c}{\begin{tabular}[c]{@{}c@{}}Calc.\\ (this work)\end{tabular}} \\
\hline 
 & 0.0 & 0.0 & 0.0 & 0.0 \\
 & 8.18 & 8.31 & 8.05 & 8.07 \\
$^4I_{9/2}$ & --- & 28.27 & 23.10 & 22.98 \\
 & 37.815 & 35.58 & 25.58 & 25.52 \\
 & --- & 57.03 & 27.69 & 27.72 \\ \hline 
 & --- & 240.78 & --- & 237.35 \\
 & --- & 247.47 & --- & 239.27 \\
$^4I_{11/2}$ & --- & --- & --- & 248.52 \\
 & --- & 263.71 & --- & 252.36 \\
 & --- & 269.42 & --- & 255.83 \\
 & --- & 276.61 & --- & 256.73 \\ \hline
 & 480.81 & 480.81 & --- & 480.51 \\
 & 490.36 & --- & --- & 483.16 \\
 & 493.33 & 490.36 & --- & 494.12 \\
$^4I_{13/2}$ & --- & --- & --- & 496.47 \\
 & 513.42 & --- & --- & 500.87 \\
 & 520.98 & 520.98 & --- & 502.45 \\
 & --- & --- & --- & 503.26 \\
\hline\hline
\end{tabular}
\caption{
\label{tab: 4}
Comparison of the experimental and calculated levels of the first three J-multiplets in meV for \bnzo.}
\end{table}

\begin{table*}
\begin{tabular}{c|cccccccccc}
\hline\hline
$m_J$ & $0.0$ & $0.0$ & $8.07$ & $8.07$ & $22.98$ & $22.98$ & $25.52$ & $25.52$ & $27.72$ & $27.72$\\
\hline
$-9/2$ &  -0.096 &  &  &  -0.373 & & 0.191 & & 0.655 &  0.621 &  \\
$-7/2$ &  &  0.519 &  -0.051 &  & 0.14 &  & 0.582 &  &  &  -0.607\\
$-5/2$ &  0.596 &  &  &  -0.075 &  & 0.631 &  & -0.401 & 0.277 &  \\
$-3/2$ &  &  -0.497 &  -0.565 &  & 0.49 &  & -0.15 &  &  &  -0.41\\
$-1/2$ &  -0.339 &  &  &  0.728 & & 0.553 & & 0.218 & -0.015 &  \\
$1/2$ &  &  -0.339 &  0.728 &  & 0.553 &  & 0.218 &  &  & -0.015\\
$3/2$ &  -0.497 &  &  &  -0.565 & & 0.49 &  & -0.15 & -0.41 &  \\
$5/2$ &  &  0.596 &  -0.075 &  & 0.631 & & -0.401 & &  &  0.277\\
$7/2$ &  0.519 &  &  &  -0.051 &  & 0.14 & & 0.582 &  -0.607 &  \\
$9/2$ &  &  -0.096 &  -0.373 &  & 0.191 &  & 0.655 & &  &  0.621\\ \hline
$-11/2$ &  &  0.002 &  -0.029 &  & -0.008 &  & -0.009 &  &  &  -0.006\\
$-9/2$ &  0.029 &  &  &  0.038 & & 0.002 & & -0.004 & 0.02 &  \\
$-7/2$ &  &  -0.04 &  0.011 &  & -0.016 &  & 0.017 & &  &  -0.012\\
$-5/2$ &  -0.028 &  &  &  -0.016 &  & -0.002 &  & -0.015 &  0.021 &  \\
$-3/2$ &  &  0.009 &  0.005 &  &  & & -0.026 &  &  & 0.001\\
$-1/2$ &  -0.008 &  &  &  -0.002 &  & -0.008 &  & 0.008 & -0.025 &  \\
$1/2$ &  &  0.008 &  0.002 &  & 0.008 &  & -0.008 & &  &  0.025\\
$3/2$ &  -0.009 &  &  &  -0.005 &  &  &  & 0.026 &  -0.001 &  \\
$5/2$ &  &  0.028 &  0.016 &  & 0.002 &  & 0.015 &  &  &  -0.021\\
$7/2$ &  0.04 &  &  &  -0.011 &  & 0.016 & & -0.017 & 0.012 &  \\
$9/2$ &  &  -0.029 &  -0.038 &  & -0.002 &  & 0.004 &  &  &  -0.02\\
$11/2$ &  -0.002 &  &  &  0.029 &  & 0.008 &  & 0.009 &  0.006 &  \\
\hline\hline
\end{tabular}
\caption{
\label{tab: 3}
Tabulated wave functions of the crystal field states in \bnzo~obtained within the intermediate coupling approximation, only the $J=9/2$ and $J=11/2$ mixing is presented. 
The crystal-field energies (in meV) are tabulated horizontally, the $m_J$-values of the ground-state multiplet vertically. Only coefficients of the 
wave functions $> 10^{-3}$ are shown.}
\end{table*}


\clearpage

\section{phonon calculations}\label{phononcalcs}

\renewcommand{\thefigure}{B\arabic{figure}}
\renewcommand{\thetable}{B\Roman{table}}
\setcounter{figure}{0}
\setcounter{table}{0}

\begin{figure}[h]
\includegraphics[width=0.98\columnwidth]
                {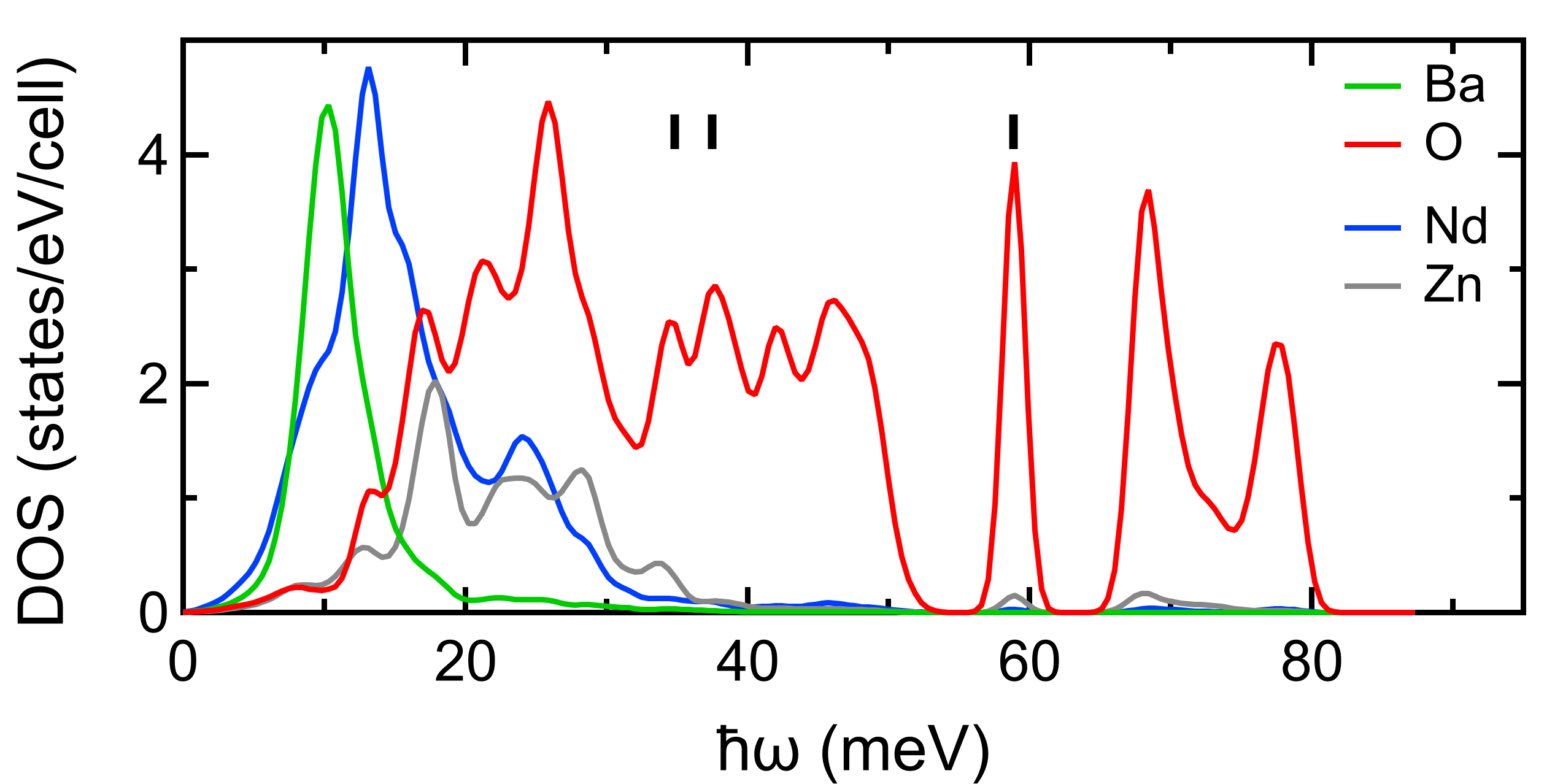}
\caption{
\label{dosfigure}
Calculated partial phonon density of states of \bnzo{} as a function of energy transfer.  Dark vertical lines are shown at 34.83, 37.51, and 58.84 meV energy transfer for comparison to the calculated oxygen partial phonon density of states (red curve).}
\end{figure}

As an additional means of checking for phonon contributions to the measured scattering intensity, the phonon DOS of \bnzo{} was calculated based on the frozen phonon method using the PHONOPY code\cite{togo_first_2015}. Forces were calculated based on density functional theory (DFT) with Perdew-Burke-Ernzerhof (PBE) exchange-correlation functional\cite{perdew} implemented in the VASP code.\cite{kresse_efficiency_1996} The interaction between ions and electrons was described by projector augmented wave method.\cite{kresse_from_1999} The 4f electrons of Nd ions are frozen in the core because the 4f states are highly localized and have little interaction with valence states. The valence wavefunctions were expanded in a plane-wave basis with a cut-off energy of 520 eV.  The calculated partial phonon density of states are shown in Fig.~\ref{dosfigure} as a function of energy transfer.  The calculated phonon branches span energy transfers between approximately 10 meV up to 80 meV.  Higher energy branches are due to several flat band optical oxygen phonon modes.  We note that three particular oxygen phonon modes are at energy transfers similar to prior observed modes in optical spectroscopy measurements as indicated by the heavy vertical lines in Fig.~\ref{dosfigure}. 

\clearpage
\section{Structural properties at 300 K}\label{struct_prop}

\renewcommand{\thefigure}{C\arabic{figure}}
\renewcommand{\thetable}{C\Roman{table}}
\setcounter{figure}{0}
\setcounter{table}{0}

The Rietveld refinement of the structural model at 300 K was done with the GSAS-II software package\cite{gsas_2}. Analysis was performed with neutron diffraction data (POWGEN\cite{powgen2011}, SNS) and synchrotron x-ray diffraction data (BL15XU, Spring-8). Two phases where used in the refinements, the primary phase \bnzo{} and an impurity phase Nd$_2$O$_3$.  The phase fraction for the Nd$_2$O$_3$ impurity phase is 1.5 \%. Table~\ref{tab: 1si} shows the final parameters of our refinement. For the neutron diffraction measurements using POWGEN, the sample was loaded into a vanadium sample can with a He exchange gas. The POWGEN measurements were conducted at 300 K using the high intensity configuration with the center of the wavelength band at 0.8 \AA. The synchrotron experiment was performed at room temperature with the precision powder X-ray diffractometer installed at the BL15XU beamline, SPring-8, Japan \cite{xray1,xray2}

\begin{figure}
\includegraphics[width=0.9\columnwidth]
                {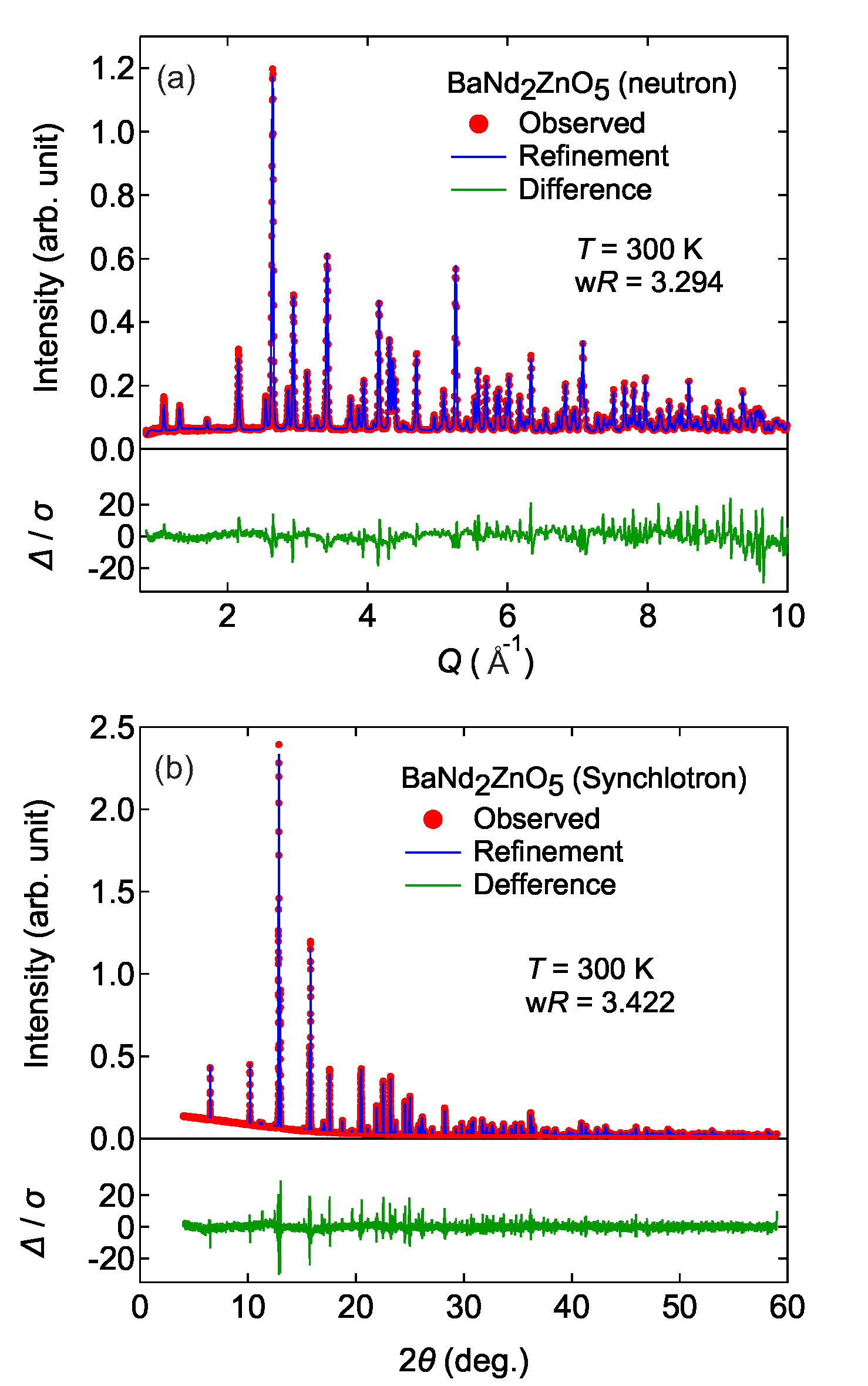}
\caption{
\label{powgen}
(a) Neutron diffraction data collected with POWGEN.  The observed data (red circles), result of refinements (blue lines), and difference between them (green).  (b) Synchrotron x-ray diffraction data collected with BL15XU. The observed data (red circles), result of refinements (blue lines), and difference between them (green).}
\end{figure}

\begin{table}[h]
\begin{tabular}{c|cccccccccc} \hline \hline
 Atom (Position) & x & y & z  \\ 
      & U$_{11}$ & U$_{22}$ & U$_{33}$ \\ 
      & U$_{12}$ & U$_{13}$ & U$_{23}$ \\ \hline
 Ba (4a) & 0 & 0 & 0.25  \\ 
     & 0.00535(12) & 0.00535(12) & 0.00659(19) \\ 
     & 0 & 0 & 0 \\ 
 Nd (8h) & 0.17399(3) & 0.67399(3) & 0 \\ 
     & 0.00320(6) & 0.00320(6) & 0.00418(9) \\ 
     & -0.00098(8) & 0 & 0 \\ 
 Zn (4b) & 0 & 0.5 & 0.25 \\ 
     & 0.00569(13) & 0.00569(13) & 0.00337(21) \\ 
     & 0 & 0 & 0 \\ 
 O1 (4c) & 0 & 0 & 0 \\ 
     & 0.00723(16) & 0.00723(16) & 0.00958(28) \\ 
     & 0 & 0 & 0 \\ 
 O2 (16l) & 0.35458(4) & 0.85458(4) & 0.13111(3) \\ 
     & 0.00894(9) & 0.00894(9) & 0.00770(12) \\ 
     & -0.00193(10) & -0.00224(7) & -0.00224(7) \\ 
     \hline \hline
\end{tabular}
\caption{
\label{tab: 1si}
Structural refinement parameters. The space group and lattice parameters are $I$4/$mcm$, $a$ = 6.76348(2) \AA \ and $c$ = 11.54204(3) \AA, respectively.} 
\end{table}

\clearpage
\newpage

\section{Magnetic space group and symmetry}\label{ap_mag_sg}

\renewcommand{\thefigure}{D\arabic{figure}}
\renewcommand{\thetable}{D\Roman{table}}
\setcounter{figure}{0}
\setcounter{table}{0}



\begin{table}[h]
\renewcommand{\arraystretch}{1.3}
\begin{tabular}{c}
\hline\hline
 Symmetry operations \\
\hline
(x,x+$\frac{1}{4}$,0$|$m$_x$,m$_y$,0) (-x,-x+$\frac{1}{4}$,0$|$m$_x$,m$_y$,0)\\
  (-x+$\frac{1}{4}$,x,0$|$-m$_y$,m$_x$,0) (x+$\frac{1}{4}$,-x,0$|$-m$_y$,m$_x$,0)\\
  (x+$\frac{1}{4}$,x,$\frac{1}{2}|$m$_y$,m$_x$,0) (-x+$\frac{1}{4}$,-x,$\frac{1}{2}|$m$_y$,m$_x$,0)\\
  (-x,x+$\frac{1}{4}$,$\frac{1}{2}|$m$_x$,-m$_y$,0) (x,-x+$\frac{1}{4}$,$\frac{1}{2}|$m$_x$,-m$_y$,0)\\
  (x,x+$\frac{3}{4}$,0$|$-m$_x$,-m$_y$,0) (-x,-x+$\frac{3}{4}$,0$|$-m$_x$,-m$_y$,0)\\
  (-x+$\frac{1}{4}$,x+$\frac{1}{2}$,0$|$m$_y$,-m$_x$,0) (x+$\frac{1}{4}$,-x+$\frac{1}{2}$,0$|$m$_y$,-m$_x$,0)\\
  (x+$\frac{1}{4}$,x+$\frac{1}{2}$,$\frac{1}{2}|$-m$_y$,-m$_x$,0) (-x+$\frac{1}{4}$,-x+$\frac{1}{2}$,$\frac{1}{2}|$-m$_y$,-m$_x$,0)\\
  (-x,x+$\frac{3}{4}$,$\frac{1}{2}|$-m$_x$,m$_y$,0) (x,-x+$\frac{3}{4}$,$\frac{1}{2}|$-m$_x$,m$_y$,0)\\
  (x+$\frac{1}{2}$,x+$\frac{1}{4}$,0$|$-m$_x$,-m$_y$,0) (-x+$\frac{1}{2}$,-x+$\frac{1}{4}$,0$|$-m$_x$,-m$_y$,0)\\
  (-x+$\frac{3}{4}$,x,0$|$m$_y$,-m$_x$,0) (x+$\frac{3}{4}$,-x,0$|$m$_y$,-m$_x$,0)\\
  (x+$\frac{3}{4}$,x,$\frac{1}{2}|$-m$_y$,-m$_x$,0) (-x+$\frac{3}{4}$,-x,$\frac{1}{2}|$-m$_y$,-m$_x$,0)\\
  (-x+$\frac{1}{2}$,x+$\frac{1}{4}$,$\frac{1}{2}|$-m$_x$,m$_y$,0) (x+$\frac{1}{2}$,-x+$\frac{1}{4}$,$\frac{1}{2}|$-m$_x$,m$_y$,0)\\
  (x+$\frac{1}{2}$,x+$\frac{3}{4}$,0$|$m$_x$,m$_y$,0) (-x+$\frac{1}{2}$,-x+$\frac{3}{4}$,0$|$m$_x$,m$_y$,0)\\
  (-x+$\frac{3}{4}$,x+$\frac{1}{2}$,0$|$-m$_y$,m$_x$,0) (x+$\frac{3}{4}$,-x+$\frac{1}{2}$,0$|$-m$_y$,m$_x$,0)\\
  (x+$\frac{3}{4}$,x+$\frac{1}{2}$,$\frac{1}{2}|$m$_y$,m$_x$,0) (-x+$\frac{3}{4}$,-x+$\frac{1}{2}$,$\frac{1}{2}|$m$_y$,m$_x$,0)\\
  (-x+$\frac{1}{2}$,x+$\frac{3}{4}$,$\frac{1}{2}|$m$_x$,-m$_y$,0) (x+$\frac{1}{2}$,-x+$\frac{3}{4}$,$\frac{1}{2}|$m$_x$,-m$_y$,0)\\
\hline\hline
\end{tabular}
\caption{
\label{magneticspacegrouptable}
The magnetic structure of \bnzo{} belongs to the magnetic space group, $P_C4/nnc$ (\#126.385) with a single Nd site (0.086, 0.336, 0.00) and lattice parameters $a_m$,$b_m$=$2a$ and $c_m$=$c$ where $a$ and $c$ are the lattice parameters of the tetragonal parent structure. The two wave-vectors defining this structure (0.5,0.5,0) and (0.5,-0.5,0) operate on different Nd dimers to form an orthogonal spin configuration of ferromagnetic dimers as depicted in Fig \ref{magstruct} in the main text. The symmetry operations presented here where generated using the Bilbao Crystallographic Server\cite{bilbao}.}
\end{table}

\end{document}